\documentclass[12pt,preprint]{aastex}

\begin{document}

\shortauthors{Allen et al.}
\shorttitle{New Disks in Lupus}

\title{New Low-Mass Stars and Brown Dwarfs with Disks in Lupus\altaffilmark{1}}

\author{
P. R. Allen\altaffilmark{2},
K. L. Luhman\altaffilmark{2},
P. C. Myers\altaffilmark{3},
S. T. Megeath\altaffilmark{4}, 
L. E. Allen\altaffilmark{3},
L. Hartmann\altaffilmark{5},
and G. G. Fazio\altaffilmark{3}}

\altaffiltext{1}
{This work is based on observations made with the {\it Spitzer Space 
Telescope}, which is operated by the Jet Propulsion Laboratory, California 
Institute of Technology under NASA contract 1407.  Support for this work 
was provided by NASA through contract 1256790 issued by JPL/Caltech. 
Support for the IRAC instrument was provided by NASA through contract
960541 issued by JPL.}
 
\altaffiltext{2}{Department of Astronomy and Astrophysics, The Pennsylvania 
State University, University Park, PA 16802; pallen@astro.psu.edu.}

\altaffiltext{3}{Harvard-Smithsonian Center for Astrophysics, 60 Garden St., 
Cambridge, MA 02138.}

\altaffiltext{4}{Department of Physics and Astronomy, The University of Toledo, 
2801 West Bancroft St., Toledo, OH 43606.}

\altaffiltext{5}{Department of Astronomy, The University of Michigan, 
500 Church Street, 830 Dennison Building, Ann Arbor, Michigan 48109.}

\begin{abstract}

Using the Infrared Array Camera and the Multiband Imaging Photometer 
aboard the {\it Spitzer Space Telescope}, we have obtained images of
the Lupus~3 star-forming cloud at 3.6, 4.5, 5.8, 8.0, and 24~\micron. 
We present photometry in these bands for the 41 previously known members
that are within our images. In addition, we have identified 19 possible
new members of the cloud based on red 3.6-8.0~\micron\ colors that are 
indicative of circumstellar disks. We have performed optical spectroscopy
on 6 of these candidates, all of which are confirmed as young low-mass
members of Lupus~3. The spectral types of these new members range from
M4.75 to M8, corresponding to masses of 0.2-0.03~$M_\odot$ for ages of 
$\sim1$~Myr according to theoretical evolutionary models. 
We also present optical spectroscopy 
of a candidate disk-bearing object in the vicinity of the Lupus~1
cloud, 2M~1541-3345, which Jayawardhana \& Ivanov recently classified as
a young brown dwarf ($M\sim0.03$~$M_\odot$) with a spectral type of M8.
In contrast to their results, we measure an earlier spectral type of 
M5.75$\pm$0.25 for this object, indicating that it is probably a low-mass star 
($M\sim0.1$~$M_\odot$).  In fact, according to its gravity-sensitive absorption
lines and its luminosity, 2M~1541-3345 is 
older than members of the Lupus clouds ($\tau\sim1$~Myr) and instead is 
probably a more evolved pre-main-sequence star that is not directly related
to the current generation of star formation in Lupus.

\end{abstract}


\section{Introduction}
\label{sec:intro}

Young embedded clusters are the primary laboratories for studies of the 
formation of stars and brown dwarfs.
The dark clouds in the constellation of Lupus are potentially attractive sites
for such studies because of their proximity to the Sun 
\citep[$d=140$-200~pc,][]{com06}.
Molecular and extinction maps have shown that the clouds extend across
$\sim20\arcdeg$ of the sky \citep{cam99,tac01} and
the richest cloud in the complex in terms of ongoing star formation
is Lupus~3. It has a fairly high surface density of young stars, 
with a peak value of $\sim200$~stars~pc$^{-2}$.
The stellar population within Lupus~3 is a good example of a
nearby, well-defined small system embedded in a filamentary dark cloud
\citep{naka03,tei05,tac06}.

Recent efforts have been made at extending the
census of the stellar population in Lupus~3 to
low-mass stars and brown dwarfs \citep{com03,lm05}.
Like their more massive counterparts, low-mass disk-bearing members
of embedded clusters can be 
identified by the presence of mid-infrared (IR) emission above 
that expected from a photosphere alone \citep{com98}. 
However, conducting a wide-field mid-IR survey for faint,
low-mass objects is technologically challenging.
For instance, the {\it Infrared Astronomical Satellite} 
was capable of mapping large fields, but it lacked the sensitivity for 
reliably reaching substellar members of nearby star-forming regions. 
With its combination of wide field of view and excellent 
sensitivity, the {\it Spitzer Space Telescope} \citep{wer04} has 
made it possible to search for low-mass stars and brown dwarfs in 
young embedded clusters like Lupus based on the presence of circumstellar 
disks \citep{all04,meg04}.

In this paper, we present the results of a search for low-mass disk-bearing
members of the most heavily populated cloud in the Lupus complex, Lupus~3,
using {\it Spitzer}'s Infrared Array Camera \citep[IRAC;][]{faz04} 
and the Multiband Imaging Photometer for {\it Spitzer} \citep[MIPS;][]{rie04}.
We begin by describing the mid-IR images of Lupus~3 obtained with these 
instruments (\S~\ref{sec:obs}) and the identification of candidate 
low-mass disk-bearing objects with these data (\S~\ref{sec:cand}). 
We then use optical spectroscopy to measure spectral types for a sample of
these candidates and to determine if they are young members of Lupus 
(\S~\ref{sec:spec}).  
In addition to the Lupus~3 targets, we include in our spectroscopic sample a 
candidate brown dwarf near Lupus~1 from \citet{all06}. 
We conclude by summarizing the properties of the new low-mass stars
and brown dwarfs found in this work (\S~\ref{sec:disc}).

\section{Infrared Images}
\label{sec:obs}

\subsection{Observations}

As a part of the Guaranteed Time Observations (GTO) of the IRAC instrument 
team, we obtained images of the Lupus 3 cloud at 3.6, 4.5, 5.8, and 
8.0~\micron\ with IRAC and at 24~\micron\ with MIPS on the {\it Spitzer Space 
Telescope}.  
The IRAC observations were performed on 2004 March 7 (UT) and
consisted of a mosaic of $6\times7$ pointings, with each pointing 
separated by $298\arcsec$ and aligned with the array axes.  
The mosaic was centered on
$\alpha=16^{\rm h}08^{\rm m}56^{\rm s}$, $\delta=-39\arcdeg08\arcmin33\arcsec$
(J2000).
Images were obtained in the high dynamic range (HDR) mode, which
consisted of one 0.4~s exposure and one 10.4~s exposure.
This mosaic was observed twice consecutively. 
The MIPS observations were performed on 2004 February 23 (UT) 
and were centered on the same position as the IRAC mosaic.
We operated MIPS in the scan mode with the medium scan rate and individual
exposure times of 3.67~s. We performed six scans, each with a length of 
$55\arcmin$.

We also make use of an IRAC map that was obtained by the Cores 2 Disks (c2d) 
Spitzer Legacy Team \citep{c2d} on 2004 September 4 (UT), which has the 
same location and dimensions as the GTO mosaic.
Because the dates of these two maps are separated by six months, 
the roll angle of the telescope differed by 180$^{\circ}$ between them. 
As a result, the combination of the two maps provides full coverage
in all four IRAC bands for the area imaged. 
Unlike the GTO mosaic, the c2d data were not taken in HDR mode. 
Instead, they consisted of two 10.4~s exposures at each pointing in a 
map that was performed once. 

\subsection{Data Reduction}

\begin{figure}
\epsscale{0.8}
\plotone{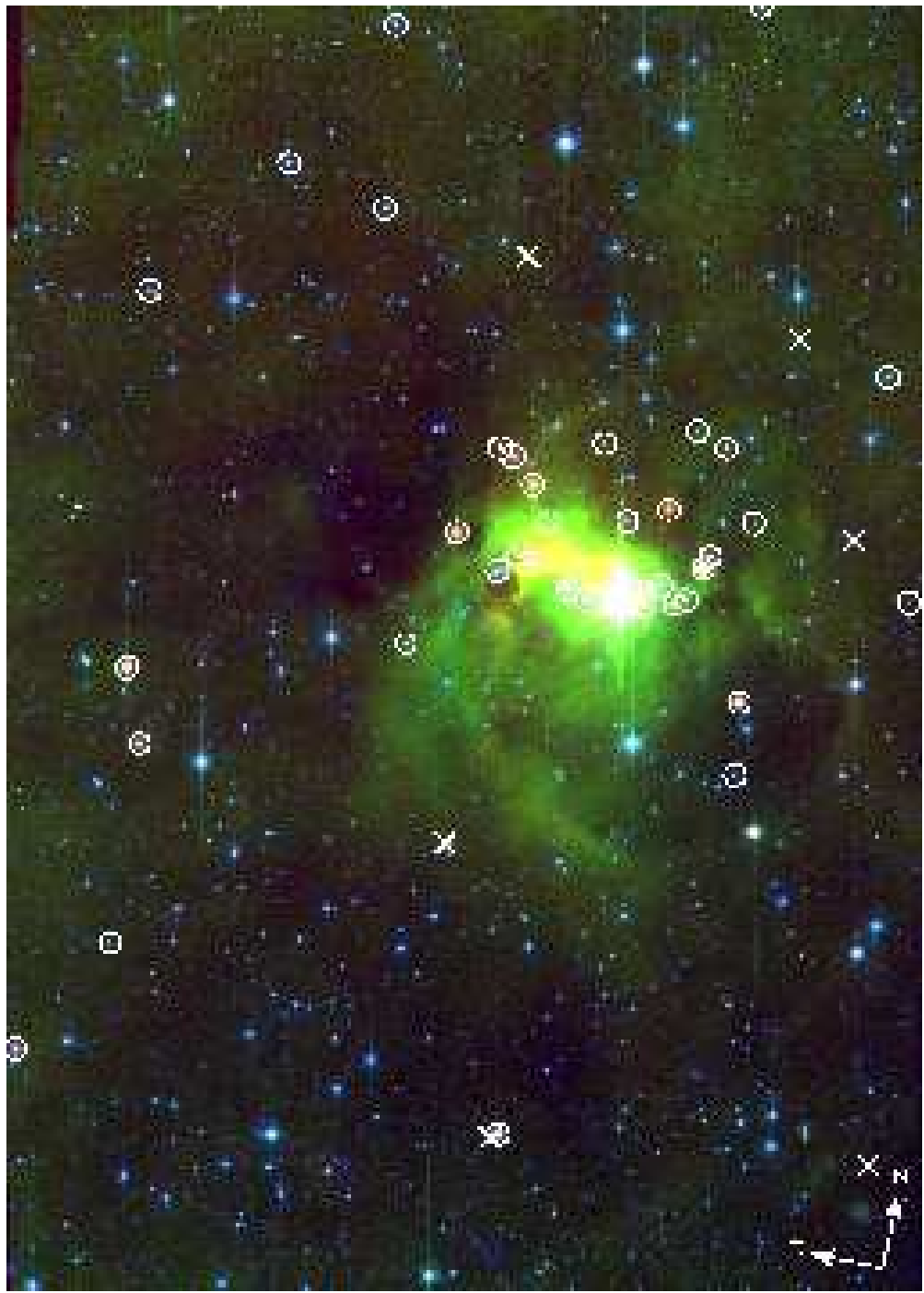}
\caption{False-color image of the Lupus 3 star-forming region consisting
of {\it Spitzer} data at 3.6, 8.0, and 24~\micron. Previously-known members
({\it circles}) and new members found in this work ({\it crosses}) are
indicated. The unmarked red source to the east of the central cluster is 
IRAS~16059-3857. The size of the image is $40\arcmin\times28\arcmin$.
}
\label{fig:image}
\end{figure}

The IRAC data were processed with the Spitzer Science Center 
(SSC) S13.2.0 pipeline, the MOPEX software from the SSC, and the IRACproc 
software \citep{iracproc}. 
For each of the four IRAC bands, we produced one 
$40\arcmin\times28\arcmin$ mosaic consisting of a combination of the GTO and 
c2d data at 10.4~s and one $35\arcmin\times28\arcmin$ mosaic consisting of the 
GTO 0.4~s data. 
The MIPS data at 24~\micron\ were reduced with MOPEX. The resulting mosaic
had dimensions of $55\arcmin\times30\farcm5$ and a total exposure time 
of 50~s at a given position.
A false-color image composed of the mosaics at 3.6, 8.0, and 24~\micron\ is 
shown in Figure~\ref{fig:image}.

We used standard IRAF photometry routines in the {\it apphot} package to 
identify all point sources in the IRAC and MIPS mosaics and to measure 
photometry for them.
For the IRAC data, we adopted zero point magnitudes ($ZP$) of 19.670, 
18.921, 16.855, 17.394 in the 3.6, 4.5, 5.8 and 8.0~\micron\ bands, where 
$M=-2.5 \log (DN/sec) + ZP$ \citep{rea05}.  For most IRAC sources, 
we measured photometry using
a radius of 4 pixels for the aperture and inner and outer radii of 
4 and 5 pixels, respectively, for the sky annulus. We used a smaller
aperture radius of 2.5 pixels for Sz~108~A and B.  The plate scale of
the reduced IRAC mosaics was $0\farcs86$~pixel$^{-1}$.
To adjust these measurements to the apertures used by \citet{rea05}, we 
applied aperture corrections of 0.18, 0.16, 0.22, and 0.48~mag for
3.6, 4.5, 5.8 and 8~$\mu$m, respectively, which we measured from bright
stars in these data.
For MIPS sources, we measured photometry with a radius of 3 pixels for
the aperture and inner and outer radii of 3 and 4 pixels, respectively, 
for the sky annulus. The plate scale of the reduced MIPS mosaic was 
$2\farcs45$~pixel$^{-1}$.
For the MIPS 24~\micron\ data we applied a flux calibration of
0.0067~mJy~(DN/sec)$^{-1}$ and an aperture correction of 0.6~mag, which 
were provided by the MIPS GTO team (David Trilling, private communication).

\section{Selection of Candidate Members of Lupus}
\label{sec:cand}

\begin{figure}
\epsscale{0.6}
\plotone{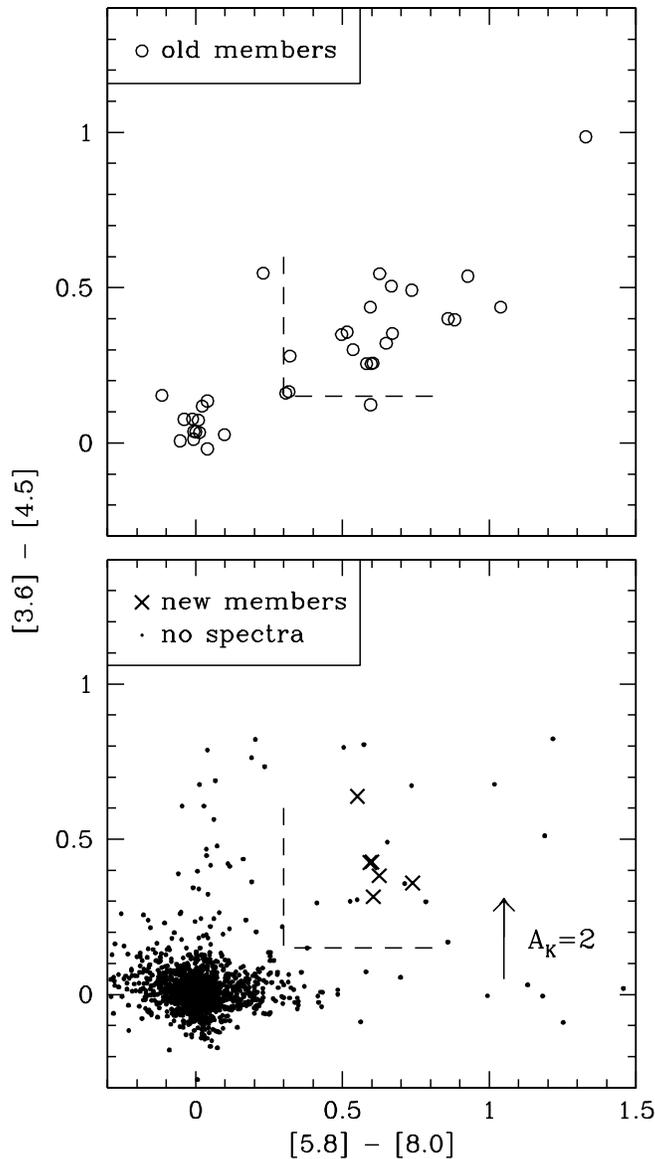}
\caption{
{\it Spitzer} IRAC color-color diagrams for the Lupus~3 star-forming region.
{\it Top}:
The previously known members of Lupus ({\it circles}, Table~\ref{tab:old})
exhibit either neutral colors consistent with stellar photospheres or 
significantly redder colors that are indicative of circumstellar disks. 
{\it Bottom}: Among the other point sources in the IRAC data, 
we selected a sample of six candidate disk-bearing members based on colors
of $[5.8]-[8.0]>0.3$ and $[3.6]-[4.5]>0.15$ ({\it dashed line}) and
obtained spectra of them (Figure~\ref{fig:spec1}),
which are all classified as new members ({\it crosses}, Table~\ref{tab:new}). 
Only objects with photometric errors less than 0.1~mag in all four bands
are shown in these two diagrams.
The reddening vector is based on the extinction law from \citet{ind05}.
}
\label{fig:1234}
\end{figure}

We can identify possible new disk-bearing members of Lupus~3 by examining
the colors of sources detected in the IRAC and MIPS images
\citep{all04,meg04,har05b,luh06tau}. To do this, we first consider the colors 
of previously known members of Lupus~3. 
From \citet{com03}, \citet{com06}, and \citet{kra97}, we have compiled
a list of 41 previously known members. Our IRAC and MIPS measurements
for these objects are given in Table~\ref{tab:old}. 
In a diagram of [3.6]-[4.5] versus [5.8]-[8.0] in Figure~\ref{fig:1234}, 
we plot the 35 previously known members that are not saturated and that have 
photometric uncertainties less than 0.1~mag in all of the IRAC bands.
The IRAC colors of the known members exhibit either neutral colors consistent
with a stellar photosphere or significantly redder colors that indicate
the presence of a circumstellar disk, which is the same behavior that has been
observed in IRAC data for young stars in other star-forming regions 
like Taurus \citep{har05b}.
As in the survey for new members of Taurus by 
\citet{luh06tau}, we adopt color criteria of 
$[3.6]-[4.5]>0.15$ and $[5.8]-[8.0]>0.3$ for selecting candidate objects
with disks in Lupus~3. We use these criteria because they encompass nearly 
all class~I and II members of Taurus \citep{har05b}. 
We applied these criteria to the $\sim1300$ 
sources with photometric errors less than 0.1~mag in all four IRAC bands, 
which are shown in Figure~\ref{fig:1234}.
This process produced 19 candidates. Most of these candidates are fainter
than the previously known members of Lupus~3, as illustrated in 
the diagram of [3.6] versus $[3.6]-[8.0]$ in Figure~\ref{fig:cmd}.
Six candidates are classified spectroscopically in \S~\ref{sec:spec}.
The IRAC and MIPS data for the 6 spectroscopic targets and the remaining
13 candidates that lack spectroscopy are listed in Tables~\ref{tab:new}
and \ref{tab:cand}, respectively. 
Five of the 19 candidates were also identified as possible members
by \citet{lm05} through optical and near-IR photometry and 7 of the
candidates were detected in X-rays by \citet{gon06}, as noted in 
Tables~\ref{tab:new} and \ref{tab:cand}.

\section{Infrared Spectroscopy of Candidates}
\label{sec:spec}

\subsection{Observations}

\begin{figure}
\epsscale{0.6}
\plotone{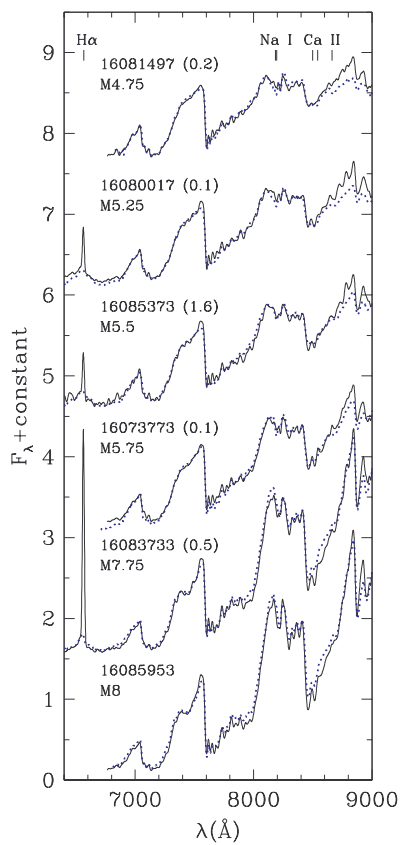}
\caption{
Optical spectra of new members of the Lupus 3 star-forming region discovered 
in this work ({\it solid lines}) compared to best-fit averages of standard 
dwarfs and giants ({\it dotted lines}). The spectra have been corrected for 
extinction, which is 
quantified in parentheses by the magnitude difference of the reddening between 
0.6 and 0.9~\micron\ ($E(0.6-0.9)$). The data are displayed at a resolution 
of 18~\AA\ and are normalized at 7500~\AA.
}
\label{fig:spec1}
\end{figure}

\begin{figure}
\epsscale{1.0}
\plotone{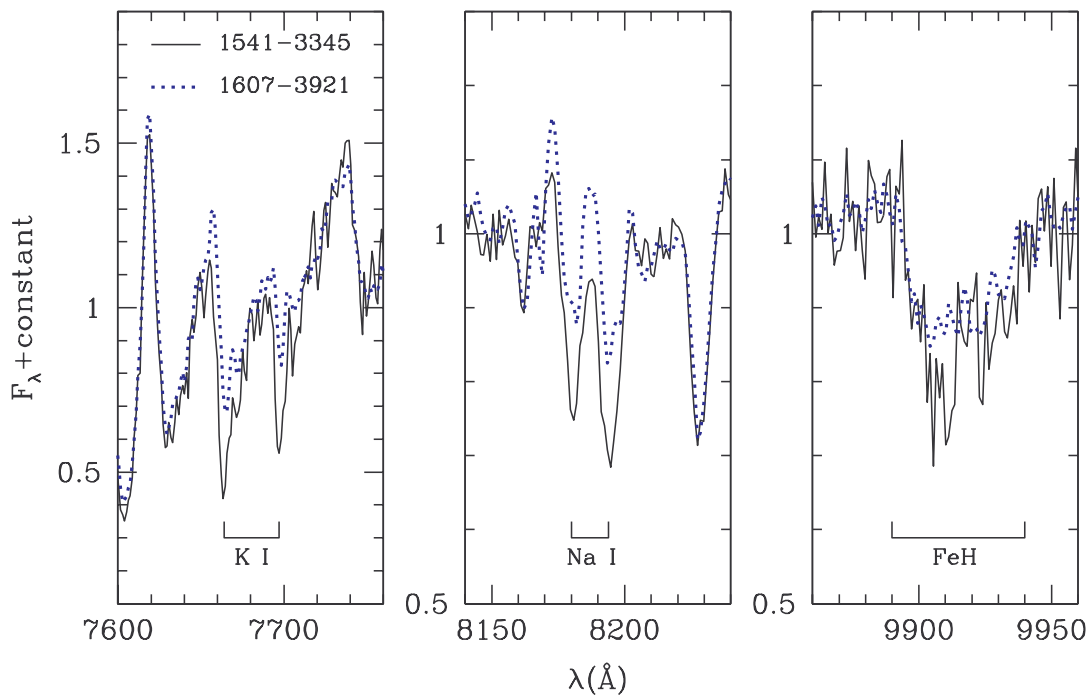}
\caption{
Gravity-sensitive absorption lines for 2M~1541-3345 ({\it solid lines})
and the M5.75 Lupus~3 member 2M~1607-3921 ($\tau\sim1$~Myr, {\it dotted 
lines}).
These data indicate that 2M~1541-3345 has a higher surface gravity than
2M~1607-3921. Thus, 2M~1541-3345 is older than $\sim1$~Myr and is not
a member of the current generation of star formation in Lupus.
The data are displayed at a resolution of 6~\AA.
}
\label{fig:spec3}
\end{figure}

We performed optical spectroscopy on 6 of the 19 candidate members of 
Lupus~3 that were identified in the previous section using the Low Dispersion 
Survey Spectrograph (LDSS-3) on the Magellan~II Telescope during the nights
of 2006 February 9 and 11.
We observed 2MASS~16081497-3857145, 16085953-3856275, and 16073773-3921388 
during the first night and 2MASS~16080017-3902595, 16083733-3923109, and
16085373-3914367 during the second night. On the first night, we also observed
an additional target, source 18 from \citet{all06}, which is also known
as 2MASS~J15414081-3345188 (hereafter 2M~1541-3345). This object was
identified as a candidate low-mass disk-bearing member of Lupus~1 in a similar
manner as our candidates \citep{all06}. The procedures for the collection and 
reduction of the optical spectra were similar to those described by 
\citet{luh04tau}. The spectra obtained on the first and second nights
exhibit wavelength coverages of 0.68-1.1 and 0.58-1~\micron\ and 
resolutions of 6 and 4~\AA, respectively.
The spectra of the six candidates in Lupus~3 are shown 
in Figure~\ref{fig:spec1} and the spectrum of 2M~1541-3345 is shown in
Figs.~\ref{fig:spec2} and \ref{fig:spec3}.

\subsection{Spectral Classification}
\label{sec:class}

To assess the membership of the candidate members of Lupus in our spectroscopic
sample and to measure their spectral types, we applied the optical 
classification methods developed in our previous studies of star-forming
regions \citep{luh99,luh03}. 
The membership diagnostics included emission lines, IR excess emission, 
gravity-sensitive spectral features (Na~I, K~I, FeH), and reddening.
We also examined the Li absorption line at 6707~\AA\ for the one spectrum that 
exhibited sufficient signal-to-noise at this wavelength.
Based on these diagnostics, we find that all of the candidates in our 
spectroscopic sample are members of the Lupus~3 cloud.  The evidence of 
membership for each object is summarized in Table~\ref{tab:new}. 
For these young objects, 
spectral types were measured through comparison to the averages of spectra 
for standard dwarfs and giants, which is the method that has been employed
for most previous optical spectral types of low-mass members of star
forming regions \citep{luh99}. 
During this process, reddening is applied to the dwarf/giant
spectrum at each spectral type to optimize the match to the target spectrum.
Thus, both spectral types and extinctions were produced. These reddening
estimates were fine-tuned by comparing each spectrum to previously known
young objects with negligible extinction \citep{luh04cha}.
The resulting spectral types for the 6 new
members are listed in Table~\ref{tab:new}. 
The extinctions are indicated in Figure~\ref{fig:spec1}, where they
are quantified by the color excess between 0.6 and 0.9~\micron\ ($E(0.6-0.9)$), 
as done by \citet{luh04cha}.  The positions of these new
members and the previously known members of Lupus~3 are indicated in 
Figure~\ref{fig:image}. 

\begin{figure}
\epsscale{0.6}
\plotone{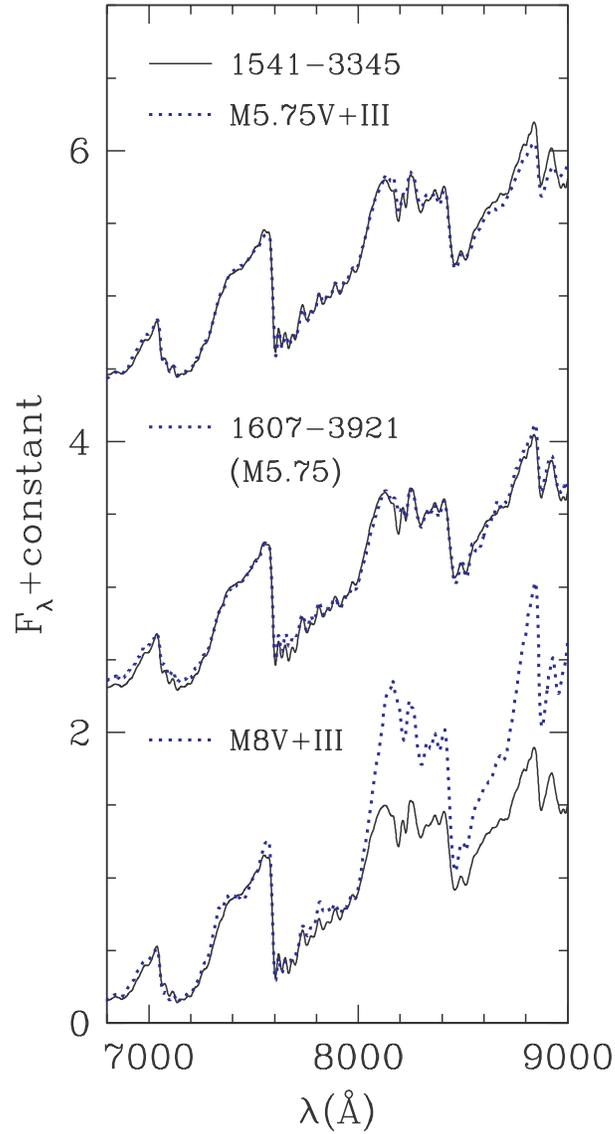}
\caption{
Optical spectrum of 2M~1541-3345 ({\it solid lines}) and three comparison
spectra ({\it dotted lines}). The spectrum of 2M~1541-3345 agrees well
with the average of a dwarf and giant at M5.75 and with the spectrum
of the M5.75 Lupus member 2M~1607-3921.
Although \citet{jay06} reported a spectral type of M8 for
2M~1541-3345, its spectrum differs significantly from that of the average of
standard M8 dwarfs and giants. The data are displayed at a resolution
of 18~\AA\ and are normalized at 7500~\AA.
}
\label{fig:spec2}
\end{figure}

As with the Lupus~3 candidates, we classified the spectrum of 2M~1541-3345
through comparison to the averages of spectra of dwarfs and giants,
arriving at a spectral type of M5.75$\pm$0.25 and an extinction of 
$A_V=0$ with an upper limit of $A_V<0.5$. Through analysis of its spectral 
energy distribution, \citet{all06} also estimated $A_V\sim0$ for 2M~1541-3345. 
The close agreement between its spectrum and the unreddened average of a 
dwarf and giant at M5.75 is demonstrated in Figure~\ref{fig:spec2}. 
In comparison, \citet{jay06} reported a spectral type of
M8 for 2M~1541-3345, but the average of M8V and M8III is a poor match to
its spectrum, as shown in Figure~\ref{fig:spec2}. 
Even if we classify our spectrum with dwarf standards as done by 
\citet{jay06} rather than averages of dwarfs and giants, we still arrive at
a spectral type of M5.75.  As with the dwarf/giant average, M8V 
is a poor match to the spectrum of 2M~1541-3345. To determine if 2M~1541-3345
is a member of the current generation of star formation in Lupus, we examined
the gravity-sensitive lines appearing in its spectrum, namely K~I,
Na~I, and FeH. We compared the strengths of these lines between 2M~1541-3345
and one of the new Lupus~3 members, 2MASS~J16073773-3921388 
(hereafter 2M~1607-3921). These two objects have the same spectral 
type, as shown in Figure~\ref{fig:spec2}, and were observed on the same night.
In Figure~\ref{fig:spec3}, we compare the K~I, Na~I,
and FeH for these two objects. These lines are significantly stronger in
2M~1541-3345 than in 2M~1607-3921, indicating that the former has a higher
surface gravity and hence greater age. Thus, we conclude that 2M~1541-3345
is not a member of the latest generation of $\sim1$~Myr-old stars that 
are arising from the Lupus clouds. Instead, 2M~1541-3345 is an older object,
possibly a member of one of the older populations in 
the Sco-Cen complex \citep{pm06}.

\section{Discussion}
\label{sec:disc}

\begin{figure}
\epsscale{1.0}
\plotone{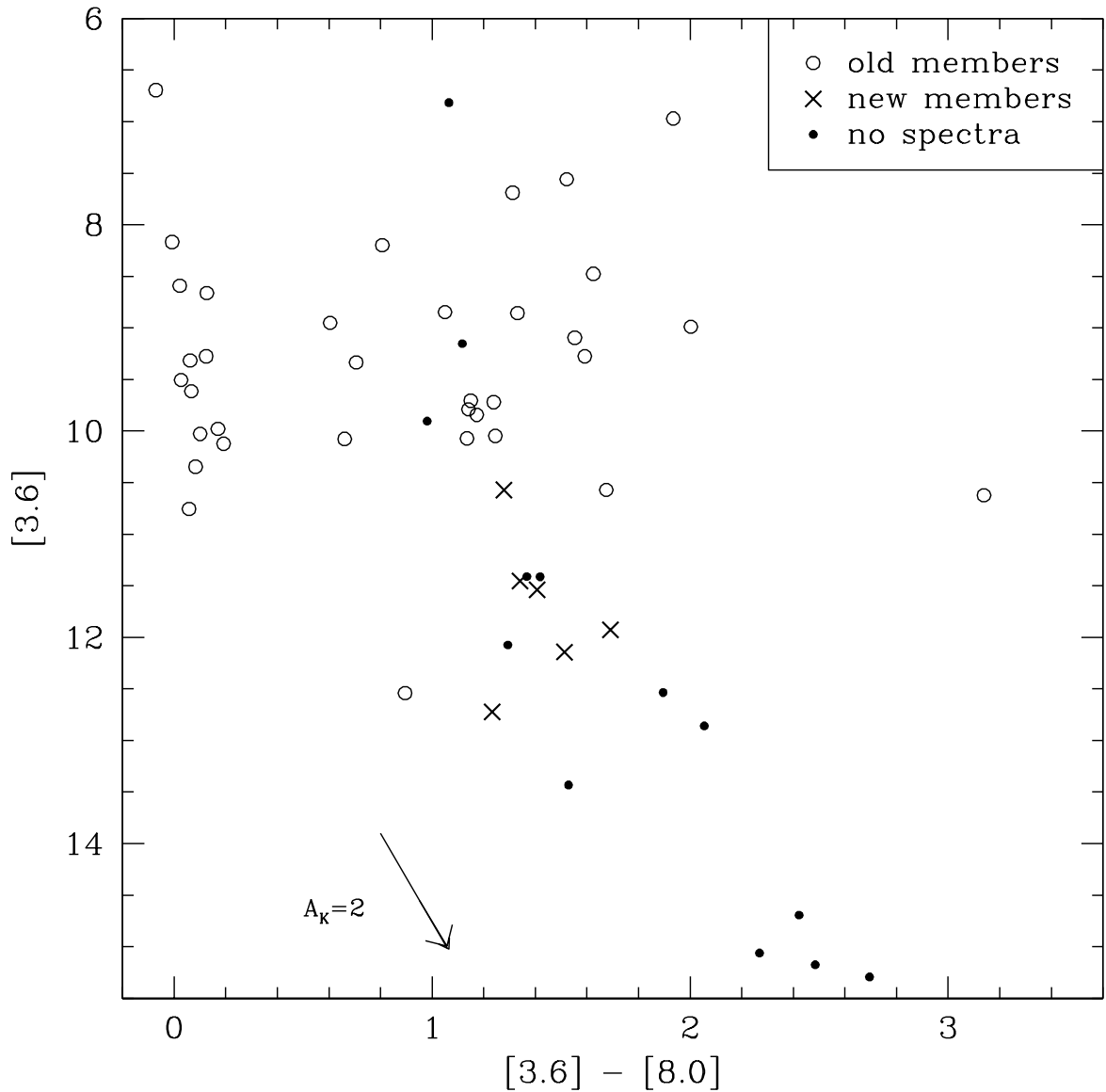}
\caption{
{\it Spitzer} IRAC color-magnitude diagram for the Lupus~3 star-forming region.
We show the previously known members of Lupus 
({\it circles}), objects that have red colors in Figure~\ref{fig:1234} and
that have been spectroscopically classified as new members 
({\it crosses}, Table~\ref{tab:new}),
and the remaining candidate disk-bearing members from Figure~\ref{fig:1234}
that lack spectroscopy ({\it filled circles}, Table~\ref{tab:cand}).
Only objects with photometric errors less than 0.1~mag in all four bands
are shown in this diagram.
The reddening vector is based on the extinction law from \citet{ind05}.
}
\label{fig:cmd}
\end{figure}

We have presented images of the Lupus~3 star-forming cloud at
3.6, 4.5, 5.8, 8.0, and 24~\micron\ obtained with IRAC and MIPS aboard
the {\it Spitzer Space Telescope}. After measuring photometry for all
point sources detected in these data, we searched the data for
new disk-bearing members of the cloud through red colors at 3.6-8.0~\micron. 
Among the 19 candidates identified in this way, 17 objects
exhibit excess emission at 24~\micron\ as well. 
Through optical spectroscopy of six of the candidates, we have confirmed 
that they are members of Lupus~3 and have measured spectral types 
of M4.75 to M8. These types correspond to masses of 0.2-0.03~$M_\odot$ 
according to theoretical evolutionary models \citep{bar98,cha00} and 
a compatible temperature scale \citep{luh03}. Two of the new members,
2MASS~J16083733-3923109 and J16085953-3856275, are likely to be substellar.
We note that the candidate 2MASS~J16085324-3914401 and the new member
2MASS~J16085373-3914367 are separated by only $6\arcsec$, and thus could 
comprise a wide binary system if the former is indeed a cluster member. 
Similarly, the candidate 2MASS~J16095628-3859518 
has a separation of only $10\arcsec$ from the known member HBC~627.
Most of the remaining 13 candidates that lack spectroscopy
have IRAC magnitudes that are indicative of low-mass
stars and brown dwarfs, as shown in Figure~\ref{fig:cmd}.  
The candidates with the reddest 
IRAC colors ($[3.6]-[4.5]>0.5$, $[5.8]-[8.0]>1$) are probably 
class~I brown dwarfs or galaxies.  Because of their youth and proximity, 
the new members that we have found and the remaining candidates (if confirmed)
are attractive targets for detailed studies of circumstellar disks and 
multiplicity near and below the hydrogen burning limit 
\citep{bur06ppv,luh06ppv}.

In addition to the candidates toward Lupus~3, we have performed spectral
classification on 2M~1541-3345, which is a possible disk-bearing object
that was identified in the vicinity of Lupus~1 by \citet{all06}. 
According to the same evolutionary models applied to the Lupus~3 members,
the spectral type of M5.75$\pm$0.25 that we have measured for this object 
implies a mass of $\sim0.1$~$M_{\odot}$. In comparison, \citet{jay06}
reported a significantly later spectral type of M8 for 2M~1541-3345 and 
thus derived a much lower mass ($M\sim0.03$~$M_{\odot}$).  In addition,
rather than being a member of Lupus~1 as assumed by \citet{jay06}, 
the gravity-sensitive lines in the spectrum of 2M~1541-3345 indicate
that it is older than members of the current generation of star formation
in Lupus~3 ($\tau\sim1$~Myr), and thus probably is not a member of the
stellar population forming in the Lupus~1 cloud. This conclusion is supported
by the fact that 2M~1541-3345 is much too faint to be a member of Lupus~1
given its spectral type of M5.75. For instance, the combination of this
spectral type and the luminosity of 2M~1541-3345 from \citet{all06}, 
which assumes the distance of Lupus~1, implies an age of 30-100~Myr 
according to the evolutionary models of \citet{bar98}. This age range is 
inconsistent with the ages of $\sim1$~Myr that are typical of star-forming 
clouds like Lupus~1. Thus, 2M~1541-3345 is probably similar to St34, which 
is an evolved disk-bearing pre-main-sequence star that happens to be projected 
against a much younger star-forming region \citep{wh05,har05a}.

\acknowledgements
P. A. and K. L. were supported by grant NAG5-11627 from the NASA Long-Term 
Space Astrophysics program. This work is based in part on the IRAC post-BCD 
processing software "IRACproc" developed by Mike Schuster, Massimo Marengo and 
Brian Patten at the Smithsonian Astrophysical Observatory.  
This work is based on observations made with the {\it Spitzer Space Telescope}, 
which is operated by the Jet Propulsion Laboratory at the California
Institute of Technology under NASA contract 1407. Support for the IRAC 
instrument was provided by NASA through contract 960541 issued by JPL.

\clearpage
\begin{deluxetable}{llllllllll}
\tabletypesize{\scriptsize}
\tablewidth{0pt}
\tablecaption{Previously Known Members of Lupus~3\label{tab:old}}
\tablehead{
\colhead{2MASS\tablenotemark{a}} &
\colhead{Other Name} &
\colhead{$J-H$\tablenotemark{a}} & 
\colhead{$H-K_s$\tablenotemark{a}} & 
\colhead{$K_s$\tablenotemark{a}} &
\colhead{[3.6]} &
\colhead{[4.5]} &
\colhead{[5.8]} &
\colhead{[8.0]} &
\colhead{[24]}
}
\startdata
   J16074959-3904287 &  Lupus 3 28 &    0.64 &    0.25 &    10.56 & 10.35$\pm$0.03 & 10.27$\pm$0.03 & 10.25$\pm$0.03 & 10.26$\pm$0.03 &         \nodata \\
   J16075230-3858059 &    Sz 95   &    0.73 &    0.27 &    10.01 &\nodata\tablenotemark{b} &  9.38$\pm$0.03 &  9.13$\pm$0.03 &\nodata\tablenotemark{b} &  6.12$\pm$0.04 \\
   J16075996-3857510 &  TTS 106  &    0.69 &    0.25 &     8.70 &  8.59$\pm$0.03 &  8.55$\pm$0.03 &  8.57$\pm$0.03 &  8.57$\pm$0.04 &  8.49$\pm$0.09 \\
   J16081096-3910459 &  F304   &    0.62 &    0.15 &     9.61 &  9.50$\pm$0.03 &  9.49$\pm$0.03 &  9.47$\pm$0.03 &  9.48$\pm$0.04 &  9.27$\pm$0.11 \\
   J16081263-3908334 &  HBC 615   &    0.78 &    0.39 &     8.96 &  7.69$\pm$0.03 &  7.25$\pm$0.03 &  6.97$\pm$0.03 &  6.38$\pm$0.03 &  4.21$\pm$0.03 \\
   J16081603-3903042 &  Par-Lup3-1   &    0.77 &    0.49 &    11.26 & 10.76$\pm$0.03 & 10.60$\pm$0.03 & 10.58$\pm$0.03 & 10.70$\pm$0.04 &         \nodata \\
   J16082180-3904214 &    Sz 97   &    0.69 &    0.34 &    10.22 &  9.84$\pm$0.03 &  9.59$\pm$0.03 &  9.25$\pm$0.03 &  8.67$\pm$0.04 &  5.86$\pm$0.04 \\
   J16082249-3904464 &   HK Lup   &    0.88 &    0.64 &     8.01 &  6.97$\pm$0.03 &  6.43$\pm$0.03 &  5.96$\pm$0.03 &  5.04$\pm$0.04 &  2.20$\pm$0.03 \\
   J16082279-3900591 &  RXJ1608.4-3900B  &    0.74 &    0.26 &    10.03 &  9.71$\pm$0.03 &  9.45$\pm$0.03 &  9.15$\pm$0.03 &  8.55$\pm$0.04 &  7.02$\pm$0.04 \\
   J16082404-3905494 &    Sz 99   &    0.72 &    0.47 &    10.74 & 10.07$\pm$0.03 &  9.71$\pm$0.03 &  9.45$\pm$0.03 &  8.94$\pm$0.03 &  6.18$\pm$0.04 \\
   J16082576-3906011 &   Sz 100   &    0.63 &    0.45 &     9.91 &  9.28$\pm$0.03 &  8.88$\pm$0.03 &  8.56$\pm$0.04 &  7.68$\pm$0.04 &  4.59$\pm$0.03 \\
   J16082778-3900406 & V1093 Sco   &    0.74 &    0.25 &     9.80 &  9.61$\pm$0.03 &  9.58$\pm$0.03 &  9.56$\pm$0.03 &  9.55$\pm$0.03 &  9.65$\pm$0.23 \\
   J16082843-3905324 &   Sz 101   &    0.68 &    0.33 &     9.39 &  8.95$\pm$0.03 &  8.79$\pm$0.03 &  8.66$\pm$0.03 &  8.35$\pm$0.04 &  6.34$\pm$0.07 \\
   J16082972-3903110 & Krautter's Star   &    0.90 &    1.07 &    12.58 & 10.62$\pm$0.03 &  9.64$\pm$0.03 &  8.81$\pm$0.03 &  7.48$\pm$0.04 &  3.29$\pm$0.03 \\
   J16083026-3906111 &  HBC 618   &    0.76 &    0.39 &    10.23 &  9.72$\pm$0.03 &  9.42$\pm$0.03 &  9.02$\pm$0.04 &  8.48$\pm$0.04 &  4.97$\pm$0.04 \\
   J16083081-3905488 &   Sz 104   &    0.67 &    0.35 &    10.65 & 10.05$\pm$0.03 &  9.73$\pm$0.03 &  9.45$\pm$0.03 &  8.80$\pm$0.03 &  5.30$\pm$0.05 \\
   J16083156-3847292 &  Lupus 3 38 &    0.75 &    0.30 &     8.62 &  8.33$\pm$0.03 &\nodata\tablenotemark{b} &  8.04$\pm$0.03 &\nodata\tablenotemark{b} &  6.17$\pm$0.04 \\
   J16083456-3905342 & V1027 Sco   &   -0.04 &    0.04 &     6.67 &  6.69$\pm$0.03 &  6.69$\pm$0.03 &  6.71$\pm$0.03 &  6.76$\pm$0.04 &  5.87$\pm$0.10 \\
   J16083578-3903479 &  Par-Lup3-2   &    0.51 &    0.39 &    10.34 & 10.03$\pm$0.03 &  9.95$\pm$0.03 &  9.89$\pm$0.03 &  9.93$\pm$0.04 &         \nodata \\
   J16083617-3923024 & V1094 Sco   &    0.84 &    0.38 &     8.66 &\nodata\tablenotemark{b} &        7.06$\pm$0.03 &\nodata\tablenotemark{b} &      6.30$\pm$0.04 &  4.19$\pm$0.03 \\
   J16083976-3906253 &   Sz 106   &    0.99 &    0.51 &    10.15 &  9.10$\pm$0.03 &  8.55$\pm$0.03 &  8.17$\pm$0.04 &  7.54$\pm$0.04 &  5.56$\pm$0.04 \\
   J16084179-3901370 &  Lupus 3 44 &    0.62 &    0.31 &    10.31 &  9.98$\pm$0.03 &  9.86$\pm$0.03 &  9.83$\pm$0.03 &  9.81$\pm$0.03 &  7.25$\pm$0.05 \\
   J16084273-3906183 &  Sz 108A   &    0.75 &    0.20 &     8.84 &  8.66$\pm$0.03 &  8.63$\pm$0.03 &  8.63$\pm$0.05 &  8.54$\pm$0.04 &         \nodata \\
   J16084273-3906183 &  Sz 108B   &    0.75 &    0.20 &     8.84 &  9.79$\pm$0.03 &  9.53$\pm$0.04 &  9.25$\pm$0.06 &  8.65$\pm$0.04 &  5.56$\pm$0.05 \\
   J16084816-3904192 &   Sz 109   &    0.58 &    0.32 &    10.50 & 10.12$\pm$0.03 &  9.99$\pm$0.03 &  9.97$\pm$0.03 &  9.93$\pm$0.04 &         \nodata \\
   J16084940-3905393 &  Par-Lup3-3   &    1.28 &    0.62 &     9.54 &  8.99$\pm$0.03 &  8.55$\pm$0.03 &  8.02$\pm$0.03 &  6.99$\pm$0.04 &  4.11$\pm$0.03 \\
   J16085143-3905304 &  Par-Lup3-4   &    1.20 &    0.96 &    13.30 & 12.54$\pm$0.03 & 11.99$\pm$0.03 & 11.88$\pm$0.04 & 11.65$\pm$0.06 &  6.23$\pm$0.07 \\
   J16085157-3903177 &   Sz 110   &    0.75 &    0.47 &     9.75 &  8.86$\pm$0.03 &  8.50$\pm$0.03 &  8.20$\pm$0.04 &  7.53$\pm$0.03 &  4.26$\pm$0.03 \\
   J16085427-3906057 & TYC 7851-426-1  &    0.53 &    0.16 &     8.21 &  8.17$\pm$0.03 &  8.19$\pm$0.03 &  8.21$\pm$0.03 &  8.18$\pm$0.04 &  7.42$\pm$0.10 \\
   J16085553-3902339 &   Sz 112   &    0.72 &    0.32 &     9.96 &  9.34$\pm$0.03 &  9.05$\pm$0.03 &  8.95$\pm$0.03 &  8.63$\pm$0.03 &  4.63$\pm$0.03 \\
   J16085780-3902227 &   Sz 113, Lup 609s  &    0.74 &    0.47 &    11.26 & 10.57$\pm$0.03 & 10.17$\pm$0.03 &  9.76$\pm$0.03 &  8.89$\pm$0.03 &  5.39$\pm$0.03 \\
   J16090185-3905124 & V908 Sco   &    0.72 &    0.38 &     9.32 &  8.48$\pm$0.03 &  7.98$\pm$0.03 &  7.59$\pm$0.03 &  6.85$\pm$0.03 &  3.44$\pm$0.03 \\
   J16090621-3908518 &   Sz 115   &    0.68 &    0.20 &    10.45 & 10.08$\pm$0.03 &  9.92$\pm$0.03 &  9.72$\pm$0.03 &  9.42$\pm$0.04 &  7.21$\pm$0.05 \\
   J16092320-3855547 & RXJ1609.4-3855   &    0.75 &    0.27 &     9.22 &  8.59$\pm$0.03 &\nodata\tablenotemark{b} &  8.92$\pm$0.03 &  8.95$\pm$0.04 &  8.38$\pm$0.10 \\
   J16092739-3850186 & RXJ1609.5-3850   &    0.63 &    0.26 &     8.57 &  8.36$\pm$0.03 &\nodata\tablenotemark{b} &  8.37$\pm$0.03 &\nodata\tablenotemark{b} &  8.21$\pm$0.07 \\
   J16093953-3855070 & V1095 Sco   &    0.62 &    0.21 &     8.00 &  7.91$\pm$0.03 &\nodata\tablenotemark{b} &  8.42$\pm$0.03 &\nodata\tablenotemark{b} &  7.85$\pm$0.06 \\
   J16094258-3919407 &  HBC 625   &    0.70 &    0.24 &     9.53 &  9.31$\pm$0.03 &  9.28$\pm$0.03 &  9.25$\pm$0.03 &  9.25$\pm$0.03 &  9.09$\pm$0.10 \\
   J16094434-3913301 &  HBC 626   &    0.83 &    0.42 &     9.42 &  8.85$\pm$0.03 &  8.50$\pm$0.03 &  8.29$\pm$0.04 &  7.80$\pm$0.04 &  4.97$\pm$0.03 \\
   J16094864-3911169 &   Sz 118, Lup 648   &    1.11 &    0.67 &     8.69 &  7.56$\pm$0.03 &  7.05$\pm$0.03 &  6.70$\pm$0.03 &  6.04$\pm$0.04 &  3.25$\pm$0.03 \\
   J16095399-3923275 &  Lupus 3 59 &    0.80 &    0.36 &     8.84 &  8.20$\pm$0.03 &  8.07$\pm$0.03 &  7.99$\pm$0.04 &  7.39$\pm$0.04 &  4.79$\pm$0.03 \\
   J16095707-3859479 &  HBC 627   &    0.73 &    0.25 &     9.42 &  9.28$\pm$0.03 &  9.20$\pm$0.03 &  9.16$\pm$0.03 &  9.15$\pm$0.03 &  \nodata \\
\enddata
\tablenotetext{a}{2MASS Point Source Catalog.}
\tablenotetext{b}{Saturated in IRAC}
\end{deluxetable}
\clearpage
\begin{deluxetable}{lllllllllll}
\rotate
\tabletypesize{\scriptsize}
\tablewidth{0pt}
\tablecaption{New Members of Lupus 3\label{tab:new}}
\tablehead{
\colhead{} &
\colhead{} &
\colhead{Membership} &
\colhead{} &
\colhead{} &
\colhead{} &
\colhead{} &
\colhead{} &
\colhead{} &
\colhead{} &
\colhead{} \\
\colhead{2MASS\tablenotemark{a}} &
\colhead{Spectral Type\tablenotemark{b}} &
\colhead{Evidence\tablenotemark{c}} &
\colhead{$J-H$\tablenotemark{a}} &
\colhead{$H-K_s$\tablenotemark{a}} &
\colhead{$K_s$\tablenotemark{a}} &
\colhead{[3.6]} &
\colhead{[4.5]} &
\colhead{[5.8]} &
\colhead{[8.0]} &
\colhead{[24]}}
\startdata
   J16073773-3921388\tablenotemark{d} &      M5.75 & e,NaK,FeH,ex &    0.67 &    0.44 &    12.13 & 11.54$\pm$0.03 & 11.18$\pm$0.03 & 10.87$\pm$0.03 & 10.13$\pm$0.04 &  7.77$\pm$0.07 \\
   J16080017-3902595\tablenotemark{e,f} &      M5.25 & Li,NaK,FeH,ex &    0.69 &    0.38 &    11.07 & 10.57$\pm$0.03 & 10.19$\pm$0.03 &  9.92$\pm$0.03 &  9.29$\pm$0.04 &         \nodata \\
   J16081497-3857145 &      M4.75 & NaK,FeH,ex &    1.01 &    1.07 &    13.13 & 11.93$\pm$0.03 & 11.29$\pm$0.03 & 10.79$\pm$0.03 & 10.24$\pm$0.04 &  7.56$\pm$0.07 \\
   J16083733-3923109\tablenotemark{g} &      M7.75 & e,NaK,FeH,ex,$A_V$ &    0.93 &    0.41 &    13.83 & 12.72$\pm$0.03 & 12.41$\pm$0.05 & 12.10$\pm$0.04 & 11.49$\pm$0.06 &  6.81$\pm$0.04 \\
   J16085373-3914367\tablenotemark{f} &       M5.5 & e,NaK,FeH,ex,$A_V$ &    1.53 &    0.92 &    12.52 & 11.45$\pm$0.03 & 11.03$\pm$0.03 & 10.71$\pm$0.03 & 10.11$\pm$0.04 &         \nodata \\
   J16085953-3856275\tablenotemark{f} &         M8 & NaK,FeH,ex &    0.61 &    0.46 &    12.84 & 12.14$\pm$0.03 & 11.72$\pm$0.03 & 11.22$\pm$0.03 & 10.63$\pm$0.04 &  7.62$\pm$0.05 \\
\enddata
\tablenotetext{a}{2MASS Point Source Catalog.}
\tablenotetext{b}{Uncertainties are $\pm0.25$ subclass.}
\tablenotetext{c}{Membership in Lupus~3 is indicated by $A_V\gtrsim1$ and
a position above the main sequence for the distance of Lupus~3 (``$A_V$"),
strong emission lines (``e"), weak absorption in Na~I, K~I, and FeH (``NaK" 
and "FeH"), strong Li absorption (``Li"), and IR excess emission (``ex").}
\tablenotetext{d}{Lup 713s.}
\tablenotetext{e}{Lup 706.}
\tablenotetext{f}{X-ray source from \citet{gon06}.}
\tablenotetext{g}{Lup 604s.}
\end{deluxetable}
\clearpage
\begin{deluxetable}{lllllllll}
\tabletypesize{\scriptsize}
\tablewidth{0pt}
\tablecaption{Candidate Members of Lupus~3\label{tab:cand}}
\tablehead{
\colhead{ID} &
\colhead{$J-H$\tablenotemark{a}} & 
\colhead{$H-K_s$\tablenotemark{a}} & 
\colhead{$K_s$\tablenotemark{a}} &
\colhead{[3.6]} &
\colhead{[4.5]} &
\colhead{[5.8]} &
\colhead{[8.0]} &
\colhead{[24]}
}
\startdata
2MASS J16075475-3915446 &    1.09 &    1.66 &    14.45 & 12.54$\pm$0.03 & 11.74$\pm$0.03 & 11.14$\pm$0.03 & 10.64$\pm$0.04 &  7.23$\pm$0.05 \\
2MASS J16080175-3912316\tablenotemark{b} &    1.34 &    0.81 &    14.25 & 13.43$\pm$0.03 & 13.07$\pm$0.04 & 12.62$\pm$0.05 & 11.90$\pm$0.08 &  8.39$\pm$0.10 \\
2MASS J16080618-3912225 &    1.50 &    0.86 &     7.67 &  6.81$\pm$0.03 &  6.51$\pm$0.03 &  6.28$\pm$0.03 &  5.75$\pm$0.03 &  4.06$\pm$0.03 \\
IRAC J16082811-3913098\tablenotemark{c} &      \nodata &      \nodata &       \nodata & 12.07$\pm$0.03 & 11.90$\pm$0.03 & 11.64$\pm$0.03 & 10.78$\pm$0.04 &  7.85$\pm$0.06 \\
IRAC J16083010-3922592 &      \nodata &      \nodata &       \nodata & 15.18$\pm$0.05 & 14.66$\pm$0.04 & 13.88$\pm$0.08 & 12.69$\pm$0.09 &  8.79$\pm$0.09 \\
IRAC J16083110-3856000\tablenotemark{b} &      \nodata &      \nodata &       \nodata & 15.06$\pm$0.05 & 14.26$\pm$0.04 & 13.37$\pm$0.06 & 12.79$\pm$0.09 &  8.99$\pm$0.12 \\
IRAC J16084679-3902074\tablenotemark{b} &      \nodata &      \nodata &       \nodata & 14.69$\pm$0.04 & 14.02$\pm$0.04 & 13.29$\pm$0.06 & 12.27$\pm$0.08 &  8.43$\pm$0.11 \\
2MASS J16084747-3905087 &    1.08 &    0.54 &    14.83 & 12.86$\pm$0.03 & 12.19$\pm$0.03 & 11.54$\pm$0.04 & 10.80$\pm$0.09 &  7.21$\pm$0.22 \\
2MASS J16085324-3914401\tablenotemark{b} &    1.04 &    0.48 &     9.80 &  9.15$\pm$0.03 &  8.85$\pm$0.03 &  8.59$\pm$0.03 &  8.04$\pm$0.04 &  5.27$\pm$0.04 \\
2MASS J16085529-3848481 &    0.59 &    0.38 &    12.02 & 11.41$\pm$0.03 & 10.92$\pm$0.03 & 10.70$\pm$0.03 & 10.04$\pm$0.04 &  7.73$\pm$0.05 \\
IRAC J16093418-3915127 &      \nodata &      \nodata &       \nodata & 15.29$\pm$0.09 & 14.47$\pm$0.06 & 13.81$\pm$0.09 & 12.60$\pm$0.06 &  9.76$\pm$0.29 \\
2MASS J16095628-3859518\tablenotemark{d} &    0.61 &    0.40 &    11.99 & 11.41$\pm$0.03 & 11.11$\pm$0.03 & 10.78$\pm$0.03 &  9.99$\pm$0.04 &  7.27$\pm$0.06 \\
2MASS J16100133-3906449 &    1.06 &    0.62 &    10.52 &  9.90$\pm$0.03 &  9.61$\pm$0.03 &  9.34$\pm$0.03 &  8.92$\pm$0.03 &  6.37$\pm$0.04 \\
\enddata
\tablenotetext{a}{2MASS Point Source Catalog.}
\tablenotetext{b}{X-ray source from \citet{gon06}.}
\tablenotetext{c}{Lup 607/707. This source is detected by 2MASS, but does
not appear in the 2MASS Point Source Catalog.}
\tablenotetext{d}{Lup 818s.}
\end{deluxetable}
\clearpage

\end{document}